\documentclass[prd,aps,showpacs,nofootinbib,preprint,tightenlines]{revtex4}

\newcommand{\comment}[1]{}

\usepackage{graphicx}
\usepackage{rotating}

\newcommand{\beq}{\begin{equation}}
\newcommand{\eeq}{\end{equation}}
\newcommand{\bqa}{\begin{eqnarray}}
\newcommand{\eqa}{\end{eqnarray}}

\begin{document}

\title{Collective modes of an Anisotropic Quark-Gluon Plasma II}

\preprint{ TUW-04-14 }

\author{Paul Romatschke}
\author{Michael Strickland}
\affiliation{Institut f\"ur Theoretische Physik, Technische Universit\"at Wien,
	Wiedner Hauptstrasse 8-10, A-1040 Vienna, Austria
     \vspace{1cm}
	}

\begin{abstract}
We continue our exploration of the collective modes of an anisotropic quark 
gluon plasma by extending our previous analysis to arbitrary Riemann sheets.  We 
demonstrate that in the presence of momentum-space anisotropies in the parton 
distribution functions there are new relevant singularities on the 
neighboring unphysical sheets.  We then show that for sufficiently strong 
anisotropies that these singularities move into the region of spacelike momentum and 
their effect can extend down to the physical sheet.  In order to demonstrate 
this explicitly we consider the polarization tensor for gluons propagating parallel 
to the anisotropy direction.  We derive analytic expressions for the gluon
structure functions in this case and then analytically continue them to 
unphysical Riemann sheets.  Using the resulting analytic continuations we 
numerically determine the position of the unphysical singularities. We then 
show that in the limit of infinite contraction of the distribution function 
along the anisotropy direction that the unphysical singularities move onto the physical 
sheet and result in real spacelike modes at large momenta for all 
``out-of-plane'' angles of propagation.
\end{abstract}
\maketitle
\newpage

\small

\section{Introduction}

The ultrarelativistic heavy ion collision experiments ongoing at the Brookhaven 
Relativistic Heavy Ion Collider (RHIC) and planned at the CERN Large Hadron 
Collider (LHC) will study the behavior of nuclear matter under extreme 
conditions.  Specifically, these experiments will explore the QCD phase diagram 
at large temperatures and small quark chemical potentials. Based on the data 
currently available from the RHIC collisions it seems that a thermalized state 
has been created during the collisions~\cite{rhicreview}.  Remarkably it seems 
that the thermalization proceeds rather rapidly in contradiction to estimates 
from leading order equilibrium perturbation theory.  However, to truly 
understand how  the plasma evolves and thermalizes one has to go beyond the 
equilibrium description.  In this paper we expand upon our previous studies of 
the collective modes of a quark gluon plasma which is (at least approximately) 
homogeneous and stationary but anisotropic in momentum space.  

These types of distribution functions are relevant because of the approximate 
longitudinal boost invariance in the central rapidity region of 
ultrarelativistic heavy ion collisions.  This implies that the initial 
distribution functions for the partons are practically delta functions in 
longitudinal momentum. Such an anisotropic quark-gluon plasma appears to be 
qualitatively different from the isotropic one since the quasiparticle 
collective modes can then be unstable 
\cite{Mrowczynski:xv,Mrowczynski:1996vh,Mrowczynski:2000ed,Birse:2003kn,
Randrup:2003cw,Romatschke:2003ms,Arnold:2003rq,Romatschke:2003yc,Mrowczynski:2004a}. The 
presence of these instabilities can dramatically influence the system's 
evolution leading, in particular, to its faster equilibration and 
isotropization.  Treating this problem in all of its generality is a daunting 
task.  In order to make progress we consider the limit of very high 
transverse temperatures at which the non-equilibrium collective behavior is 
describable in terms of processes in which all loop momenta are {\it hard}.  
In the case of thermal 
equilibrium {\it hard} corresponds to momenta of order $T$ but in the 
non-equilibrium case {\it hard} corresponds to an unspecified scale contained in the 
distribution function.

In a previous paper \cite{Romatschke:2003ms} we calculated the hard-loop gluon 
polarization tensor in the case that the momentum space anisotropy is obtained 
from an isotropic distribution by the rescaling of one direction in momentum 
space.  The resulting expression for the gluon polarization tensor was then decomposed 
into a four-component tensor basis and the structure functions associated with 
this tensor basis where computed numerically for general anisotropies and 
analytically in the limit of small anisotropies.  We demonstrated that a 
contraction of an isotropic distribution function along the anisotropy 
direction, $\hat{\bf n}$, resulted in one additional stable quasiparticle mode, 
two damped quasiparticle modes in the lower half plane, and two unstable 
(anti-damped) quasiparticle modes in the upper half plane.  In the case that the 
isotropic distribution was stretched along the anisotropy direction we found 
that again an additional stable quasiparticle mode was generated but only one 
damped and one anti-damped quasiparticle mode were found in this case.  The 
unstable modes found correspond to electric or magnetic type instabilities with 
the latter being analogous to the Weibel instability in QED 
plasmas~\cite{Weibel:1957,Bernstein:1957,Berger:1972,Davidson:1972,Yang:1993}.

In this paper we continue our exploration of the collective modes of an anisotropic quark 
gluon plasma by extending our previous analysis to arbitrary Riemann sheets.  We 
demonstrate that in the presence of momentum-space anisotropies in the parton 
distribution functions there are new relevant singularities on the 
neighboring unphysical sheets.  We then show that for sufficiently strong 
anisotropies that these singularities move into the region of spacelike momentum and 
their effect can extend down to the physical sheet.  In order to demonstrate 
this explicitly we consider the polarization tensor for gluons propagating parallel 
to the anisotropy direction.  We derive analytic expressions for the gluon
structure functions in this case and then analytically continue them to 
unphysical Riemann sheets.  Using the resulting analytic continuations we 
numerically determine the position of the unphysical singularities. We then 
show that in the limit of infinite contraction of the distribution function 
along the anisotropy direction that the unphysical singularities move onto the physical 
sheet and result in real spacelike modes at large momenta for all 
``out-of-plane'' angles of propagation.

The organization of the paper is as follows: In Sec.~\ref{structure:sec} we 
first review the necessary integral expressions for the hard-loop gluon 
polarization tensor and then in Sec.~\ref{kparn:sec} we present analytic 
expressions for the hard-loop gluon polarization tensor structure functions in 
the case that the gluon is propagating parallel to the anisotropy direction.  In 
Secs.~\ref{extension:sec}-\ref{almostlx:sec} we extend these expressions to 
arbitrary Riemann sheets and solve the dispersion relations for the 
singularities existing on the neighboring unphysical sheets.  In 
Sec.~\ref{lx:sec} we present analytic expressions for the gluon polarization 
tensor structure functions in the large-anisotropy limit and solve the resulting 
dispersion relations for arbitrary angle of propagation.  In Sec.~\ref{conc:sec} 
we summarize the results and speculate about the possible impact of the now 
relevant unphysical singularities.

\section{Gluon Polarization Tensor Revisited}
\label{structure:sec}

The hard-loop gluon polarization tensor of an anisotropic system is given by 
\cite{Mrowczynski:2000ed,Romatschke:2003ms}
\beq
\Pi^{i j}(K) = - 2 \pi \alpha_s%
 \int \frac{d^3 p}{(2\pi)^3} v^{i} \partial^{l} f({\bf p})
\left( \delta^{j l}+\frac{v^{j} k^{l}}{K\cdot V + i \epsilon}\right) \; ,
\label{selfenergy2}
\eeq
\comment{m}
where $K=(\omega,{\bf k})$, $V=(1,{\bf v})$, ${\bf v}={\bf p}/|{\bf p}|$ and
\beq
f({\bf p}) \equiv 2 N_f \left(n({\bf p}) + \bar n ({\bf p})\right) + 4 N_c n_g({\bf p}) \; .
\eeq
\comment{m}
Note that in (\ref{selfenergy2}) we have specialized to spacelike Lorentz 
indices; however, it is possible to derive the polarization tensor also
for arbitrary Lorentz indices.

To simplify the calculation we follow Ref.~\cite{Romatschke:2003ms} and require the
distribution function $f({\bf p})$ to be given by
\beq
f({\bf p})=f_{\xi}({\bf p}) = N(\xi) \ %
f_{\rm iso}\left(\sqrt{{\bf p}^2+\xi({\bf p}\cdot{\bf \hat n})^2}\right)%
\; .
\label{squashing}
\eeq
\comment{pm}
Here $f_{\rm iso}$ is an arbitrary isotropic distribution function, 
${\bf \hat n}$ is the direction of the anisotropy, $\xi>-1$ is a
parameter reflecting the strength of the anisotropy, and $N(\xi)$ is a
normalization constant. To fix $N(\xi)$ we require that the number density
to be the same both for isotropic and arbitrary anisotropic systems,
\beq
\int_{\bf p} f_{\rm iso}(p)=%
\int_{\bf p} f_{\xi}({\bf p})=N(\xi)%
\int_{\bf p} %
f_{\rm iso}\left(\sqrt{{\bf p}^2+\xi({\bf p}\cdot{\bf \hat n})^2}\right) ,
\eeq
\comment{pm}
and can be evaluated to be
\beq
N(\xi)=\sqrt{1+\xi} \; .
\eeq
\comment{pm}
Using an appropriate tensor basis \cite{Romatschke:2003ms}
one can then decompose the self-energy into four structure functions
$\alpha,\beta,\gamma,$ and $\delta$ by taking the contractions
\begin{eqnarray}
k^{i} \Pi^{ij} k^{j} & = & k^2 \beta \; , \nonumber \\
\tilde{n}^{i} \Pi^{ij} k^{j} & = & \tilde{n}^2 k^2 \delta \; , \nonumber \\
\tilde{n}^{i} \Pi^{ij} \tilde{n}^{j} & = & \tilde{n}^2 (\alpha+\gamma) \; , %
\nonumber \\
{\rm Tr}\,{\Pi^{ij}} & = & 2\alpha +\beta +\gamma \; ,
\label{contractions}
\end{eqnarray}
\comment{m}
where $\tilde{n}^i=(\delta^{ij} - k^i k^j/k^2) \hat{n}^j$.
The structure functions then depend on $\omega, k$ and the angle
${\bf \hat{k}} \cdot {\bf \hat{n}}=\cos{\theta_n}$ 
as well as on the strength of the anisotropy, $\xi$.  Integral expressions
for $\alpha,\beta,\gamma,$ and $\delta$ for arbitrary angle of propagation
and anisotropy parameter $\xi$ can be found in 
Ref.~\cite{Romatschke:2003ms}.\footnote{
In Ref.~\cite{Romatschke:2003ms} $N(\xi)$ 
was fixed to be $N(\xi)=1$ so the reader should make sure
to adjust for the difference where appropriate.
}

\section{Special case I :  ${\bf k} \mid\mid \hat{\bf n}$}
\label{kparn:sec}

Let us now consider the case where the momentum ${\bf k}$ is in the direction
of the anisotropy, ${\bf \hat{n}}$, i.e. $\theta_n=0$.  Using the changes of 
variables
\begin{equation}
\tilde{p}^2=p^2\left(1+\xi ({\bf v}\cdot{\bf \hat{n}})^2\right) \; ,
\label{variablechange}
\end{equation}
\comment{pm}
allows us to simplify Eq.~(\ref{selfenergy2}) to
\begin{equation}
\Pi^{i j}(K) = m_{D}^2 \sqrt{1+\xi}\int \frac{d \Omega}{4 \pi} v^{i}%
\frac{v^{l}+\xi({\bf v}.{\bf \hat{n}}) \hat{n}^{l}}{%
(1+\xi({\bf v}.{\bf \hat{n}})^2)^2}
\left( \delta^{j l}+\frac{v^{j} k^{l}}{K\cdot V + i \epsilon}\right) \; ,
\label{pikx0}
\end{equation}
\comment{m}
where 
\beq
m_D^2 = -{\alpha_s\over \pi} \int_0^\infty d p \,  
  p^2 {d f_{\rm iso}(p^2) \over dp} \; .
\eeq
\comment{m}
Taking the contractions in Eq.~(\ref{contractions}) the structure functions
can further be simplified using
\beq
{\bf k} \cdot {\bf v}= k \, {\bf \hat{n}} \cdot {\bf v} = k \cos{\theta} \; .
\eeq
\comment{m}
While one integration becomes straightforward, the remaining integration can be
performed after a little bit of algebra and one obtains for the relevant 
contractions
\bqa
\frac{\alpha}{m_D^2}&=&\frac{\sqrt{1+\xi}}{4 \sqrt{\xi} (1+\xi z^2)^2} %
\left[\left(1+z^2+\xi (-1+(6+\xi) z^2-(1-\xi) z^4)\right) %
\arctan \sqrt{\xi}\right.\nonumber \\
&& \hspace{4cm} \left.+ \sqrt{\xi} \left((z^2-1) (1+\xi z^2-(1+\xi) z \ln{\frac{%
z+1+i \epsilon}{z-1+i \epsilon}}\right)\right] \, , \nonumber \\
\frac{\beta}{m_D^2} &=& - \frac{z^2 \sqrt{1+\xi}}{2 \sqrt{\xi} (1+\xi z^2)^2} %
\left[(1+\xi)(1-\xi z^2)\arctan{\sqrt{\xi}} \right. \nonumber \\
&& \hspace{4cm} \left. + \sqrt{\xi} \left(%
(1+\xi z^2)-(1+\xi) z \ln{\frac{z+1+i \epsilon}{z-1+i \epsilon}}
\right)\right] \, , \nonumber \\
\frac{\hat{\delta}}{m_D^2} &=& \frac{z \sqrt{1+\xi}}{4 \sqrt{\xi} (1+\xi z^2)^3}%
\left[z \left(-1+\xi (3+6 \xi -2(3+6 \xi + \xi^2) z^2+\xi(3+\xi) z^4%
)\right) \arctan{\sqrt{\xi}}\right.\nonumber \\
&&\hspace{-4mm}\left.+\sqrt{\xi} \left(z(1+\xi z^2)(1+4 \xi-3 \xi z^2)+\xi%
(-1+z^2)(-1+4 z^2+3 \xi z^2) \ln{\frac{z+1+i \epsilon}{z-1+%
i \epsilon}}\right)\right]\, ,
\label{kx0strufunc}
\eqa
\comment{pm}
where $z=\omega/k$ and $\hat{\delta}=\delta k$.  Note that for this angle of
propagation the structure function $\gamma$ vanishes.

\subsection{Extension to unphysical sheets}
\label{extension:sec}

The above structure functions all possess a logarithmic cut running along the 
real $z$ axis for $z^2<1$. It is, however, possible to extend their definition 
beyond this cut, which then corresponds to an unphysical Riemann sheet of $z$. 
More precisely, there are two such unphysical sheets which can be accessed by 
either extending the physical sheet from above or below the cut.  In order to 
make this extension we first compute the structure functions on the physical 
sheet using the integral representation given in Eq.~(\ref{pikx0}). We can then 
deform the integration contour which runs from $\cos{\theta}=-1$ along the real 
axis to $\cos{\theta}=1$ into the lower half plane (LHP) and then move the 
point we are interested in from the upper to the lower half plane. Deforming the 
original contour so that it again runs along the real axis we see that we pick 
up an extra contribution corresponding to the residue at the point in the LHP.  

This procedure can be used to analytically continue the structure functions for 
all values of $\theta_n$ but in the case that $\theta_n=0$ the residue can be 
evaluated straightforwardly and is simply $- 2 \pi  i$.  The structure functions 
can likewise be extended into the upper half plane (UHP) and the residue is then 
$+2\pi i$. The resulting rule is then the expected one, namely, that the 
structure functions can be extended to unphysical sheets by the usual extension 
of the logarithm 
\bqa 
\ln\left({z+1\over z-1}\right) = 
\ln\left(\left|{z+1\over z-1}\right|\right) + i \left[ {\rm arg}\left({z+1\over 
z-1}\right) + 2 \pi n \right] \, , 
\eqa 
\comment{m} 
where $n$ specifies the sheet number.  Note that $n=1$ extends the physical 
logarithm into the UHP and $n=-1$ extends it into the LHP.  Higher $n$ 
correspond to higher sheets which can be safely ignored as we will discuss 
below.
  
\subsection{Collective modes}

The dispersion relations for the gluonic modes in an anisotropic quark-gluon
plasma are in general given by the zeros of
\bqa
\Delta_A^{-1}=k^2-\omega^2+\alpha&=&0 \; , \nonumber \\
\Delta_G^{-1}=%
(k^2-\omega^2+\alpha+\gamma)(\beta-\omega^2)-k^2 \tilde{n}^2 \delta^2&=&0 \; .
\label{disprel}
\eqa
\comment{m}
In the case ${\bf k} || {\bf \hat{n}}$, however, $\gamma$ vanishes identically,
as does $\tilde{n}^2=1-(\bf{\hat{k}}\cdot {\bf \hat{n}})^2$. Therefore,
it is sufficient to solve the equations
\bqa
k^2-\omega^2+\alpha&=&0 \; , \nonumber\\
\beta-\omega^2&=&0 \; ,
\label{tn0disprel}
\eqa
\comment{m}
which will be referred to as $\alpha$- and $\beta$-modes, respectively.  These 
modes correspond to poles in the propagator for $\alpha$- and $\beta$-modes. We 
will use this term to describe solutions on both physical and unphysical Riemann 
sheets; however, whenever we are speaking about the unphysical singularities we will 
always explicitly label them as {\em unphysical} $\alpha$- and $\beta$-modes. 
The reader should be aware that these unphysical singularities do not correspond to 
real degrees of freedom unless the solution associated with them moves onto 
the physical Riemann sheet.

Before we present the dispersion relations we would like to first count
the number of modes on the various sheets in order to be assured that we
have indeed found all solutions.
The number of modes can be counted by doing a so-called 
Nyquist analysis, based on the special case of Cauchy's integral,
\beq
N-P=\frac{1}{2\pi i} \oint_{C} dz \frac{f'(z)}{f(z)},
\label{Nyquist}
\eeq
\comment{m}
where $N$ and $P$ are the number of zeros and poles of $f(z)$ times
their multiplicity in the region encircled by the closed path $C$.

\begin{figure}[htb]
\hfill
\begin{minipage}[t]{.42\linewidth}
\includegraphics[width=6cm]{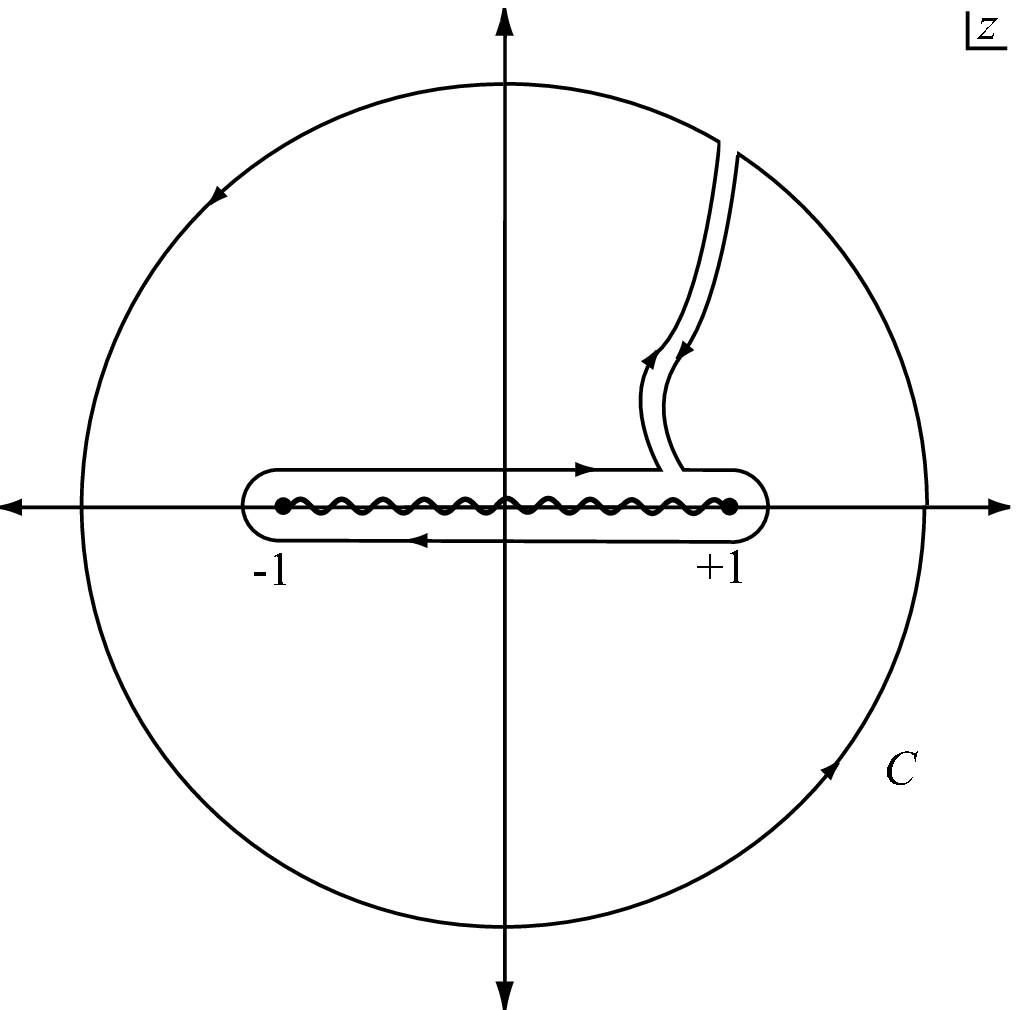}
\caption{Contour $C$ in the complex $z$ plane used for finite $\xi$ Nyquist analysis.}
\label{nyquistcontour}
\end{minipage} \hfill
\hfill
\begin{minipage}[t]{.42\linewidth}
\includegraphics[width=6cm]{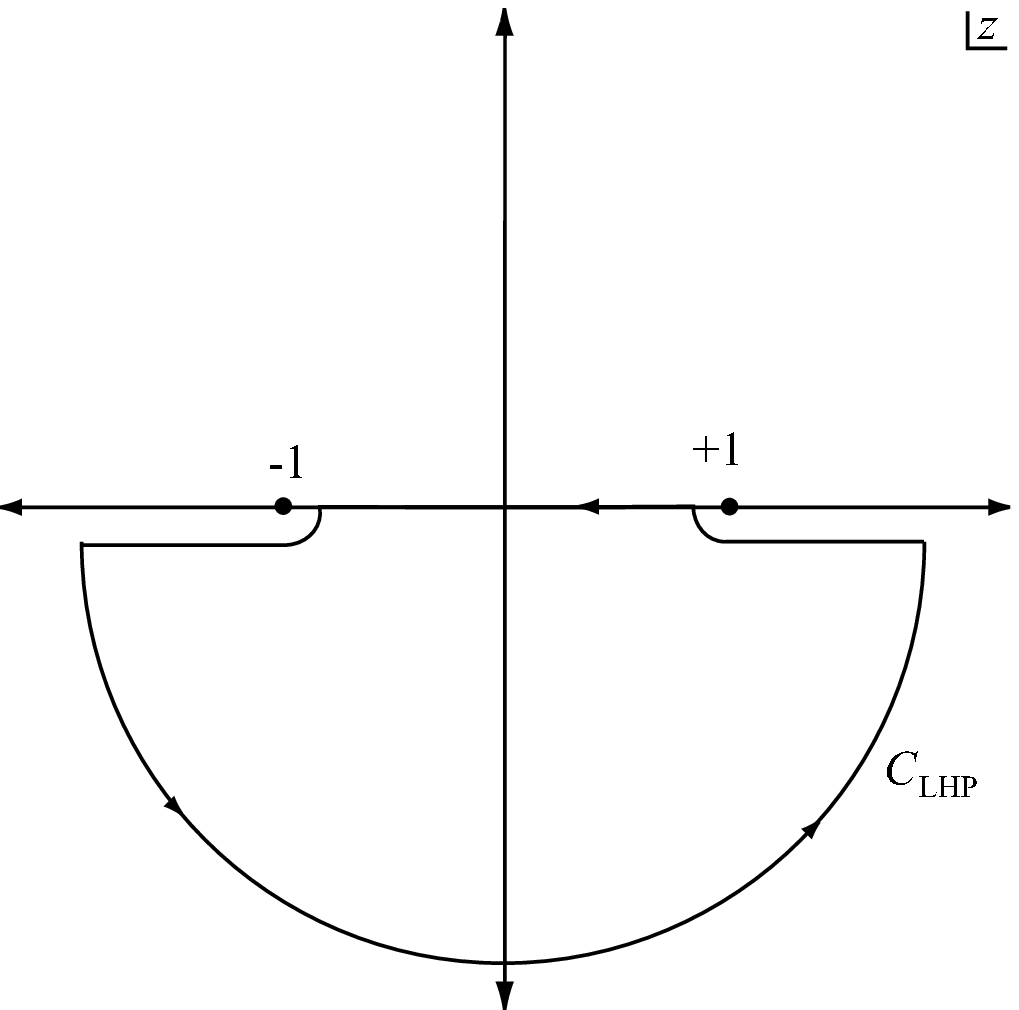}
\caption{Contour $C_{\rm LHP}$ in the complex $z$ plane used for finite $\xi$ 
Nyquist analysis on the LHP unphysical sheet.}
\label{nyquistcontour2}\end{minipage} \hfill
\end{figure}

Choosing $f(z)=k^2(1-z^2)+\alpha(z)$ and using the explicit form
of the structure function given in Eq.~(\ref{kx0strufunc}), one finds
that $f(z)$ has a logarithmic cut for real $z^2<1$, while being analytic
for all other finite $z$. One can then choose the contour $C$ depicted
in Fig.~\ref{nyquistcontour}, so that $P=0$.  We then evaluate
the respective pieces of the contour $C$, finding
\beq
N_{\alpha,phys}=\frac{1}{2 \pi i} \left(4 \pi i+0+0+4 \pi i %
\Theta(-\lim_{z\rightarrow 0}f(z))\right)=2+2%
\Theta(-\lim_{z\rightarrow 0}f(z)) \; .
\label{Nalphaphys}
\eeq
\comment{m}
The first contribution comes from the large circle at $|z|\gg 1$, while 
the first zero is the contribution from the path connecting 
the large circle with the contour around $z^2<1$; the second zero is the
contribution from the small half-circles around $z=\pm 1$.
The last contribution comes from the straight lines running infinitesimally
above and below the cut at $z^2<1$. They can be evaluated by using
\beq
\int_{-1+i \epsilon}^{1+i \epsilon} dz \frac{f'(z)}{f(z)}=%
\ln{\frac{f(1+i\epsilon)}{f(-1+i \epsilon)}} \; ,
\eeq
\comment{m}
and therefore represent the ``winding number'' of the image of $f(z)$
around the origin. Since from Eq.~(\ref{kx0strufunc}) it is clear that 
${\rm Re}\,f(-1+i \epsilon)>0$ and for $z^2<1$ one has ${\rm Im}\,f(z)=0$ 
only for $z=0$, the winding number can either be zero or one, depending on the 
sign of $\lim_{z\rightarrow 0}{\rm Re}\,f(z+i\epsilon)$. For the path with 
${\rm Im}\,z<0$ one proceeds similarly, finding the result given in 
Eq.~(\ref{Nalphaphys}).

For the $\beta$-mode, the same techniques can be applied to find the result
\beq
N_{\beta,phys}=2 \; .
\eeq
\comment{m}
These two modes, together with two modes coming from $N_{\alpha,phys}$,
correspond to the propagating and stable modes for positive and negative
(real) frequencies, while the modes that depend on the sign of the static limit
of $f(z)$ correspond to solutions with purely imaginary $z$ which are
the damed and anti-damped physical modes 
already found in Ref.~\cite{Romatschke:2003ms}.

We now want to extend the above analysis to the unphysical sheets that can be 
accessed by extending the structure functions through the cut to the LHP and
UHP, as discussed above. Let us first discuss the $\alpha$-mode in the 
unphysical LHP by introducing $f_{LHP}(z)=k^2(1-z^2)+\alpha_{LHP}(z)$ and
choosing a contour $C_{LHP}$ which is 
shown in Fig.~\ref{nyquistcontour2}. One notable difference
to the physical sheet is that now $f_{LHP}(z)$ has a pole of second order
at $z=-i/\sqrt{\xi}$, so that $P_{\alpha,LHP}=2$. Other than that the
analysis is conducted as in the case above, finding
\beq
N_{\alpha,LHP}-P_{\alpha,LHP}=-\Theta(-\lim_{z\rightarrow 0} f_{LHP}(z)) \; ,
\eeq
\comment{m}
so that $N_{\alpha,LHP}=1+\Theta(\lim_{z\rightarrow 0}f_{LHP}(z))$. 

For the UHP, one uses a similar analysis to find 
\beq
N_{\alpha,UHP}=1+\Theta(\lim_{z\rightarrow 0}f_{UHP}(z)) \; ,
\eeq
\comment{m}
while for the unphysical $\beta$-mode one has $N_{\beta,LHP}=N_{\beta,UHP}=2$. 
Closer inspection of these modes shows that the unphysical $\alpha$-modes 
correspond to solutions with purely imaginary $z$, while the unphysical $\beta$-modes 
are solutions with complex $z$ with positive/negative real and imaginary 
parts.

\begin{figure}
\includegraphics[width=0.3\linewidth]{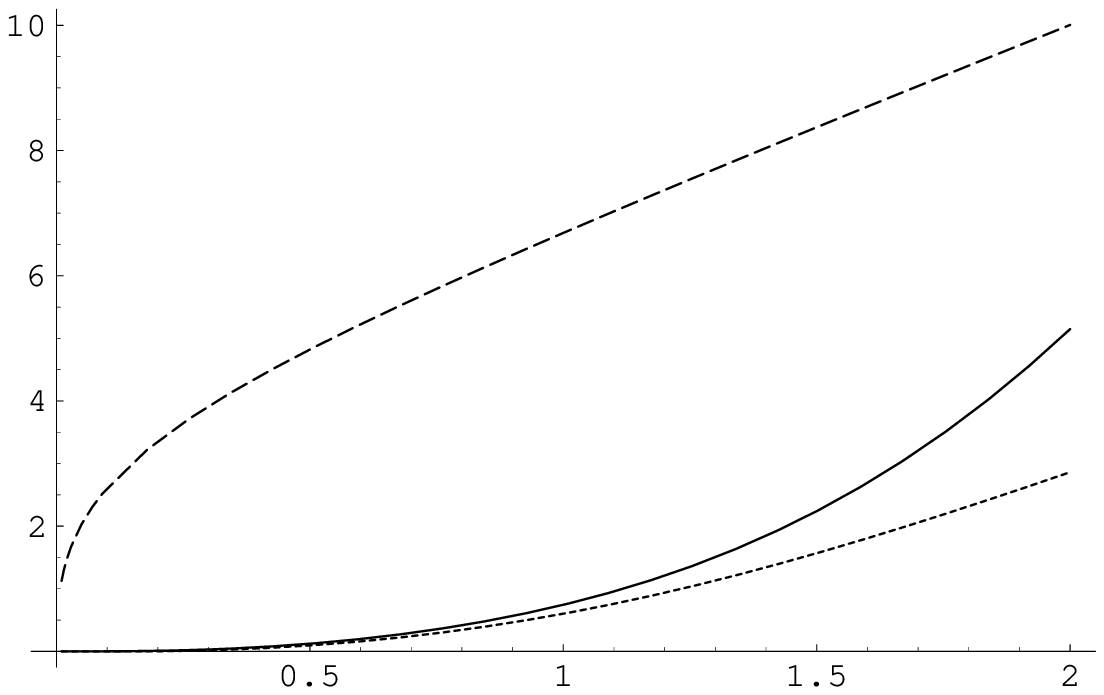}
\hspace*{0.5cm}
\includegraphics[width=0.3\linewidth]{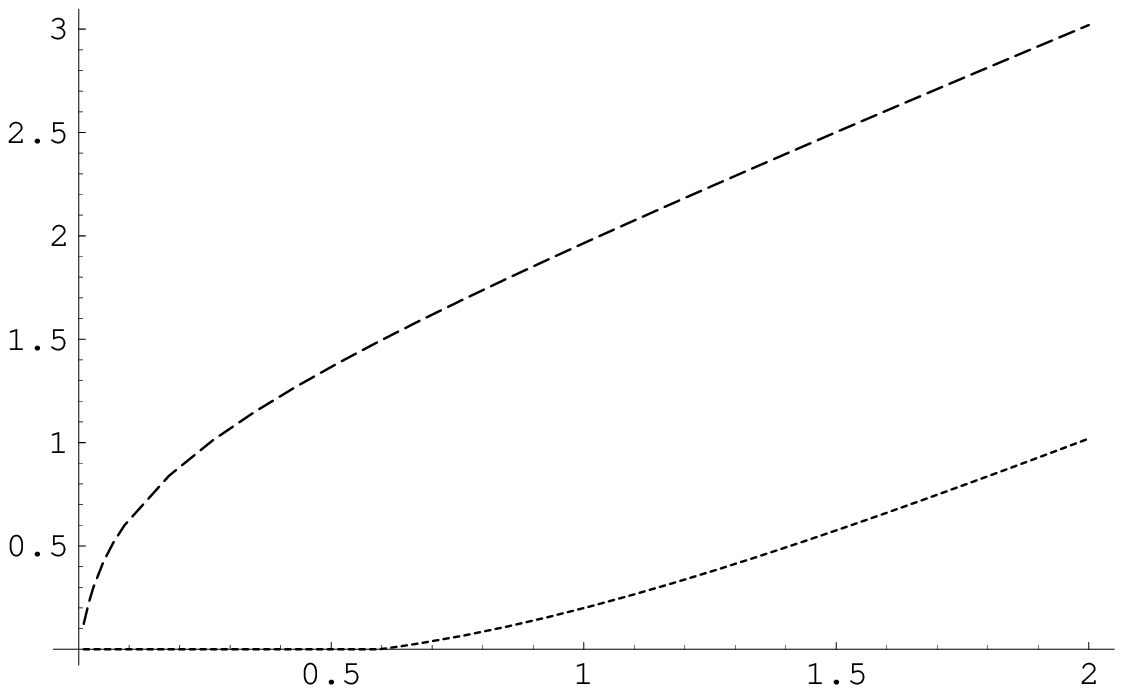}
\hspace*{0.5cm}
\includegraphics[width=0.3\linewidth]{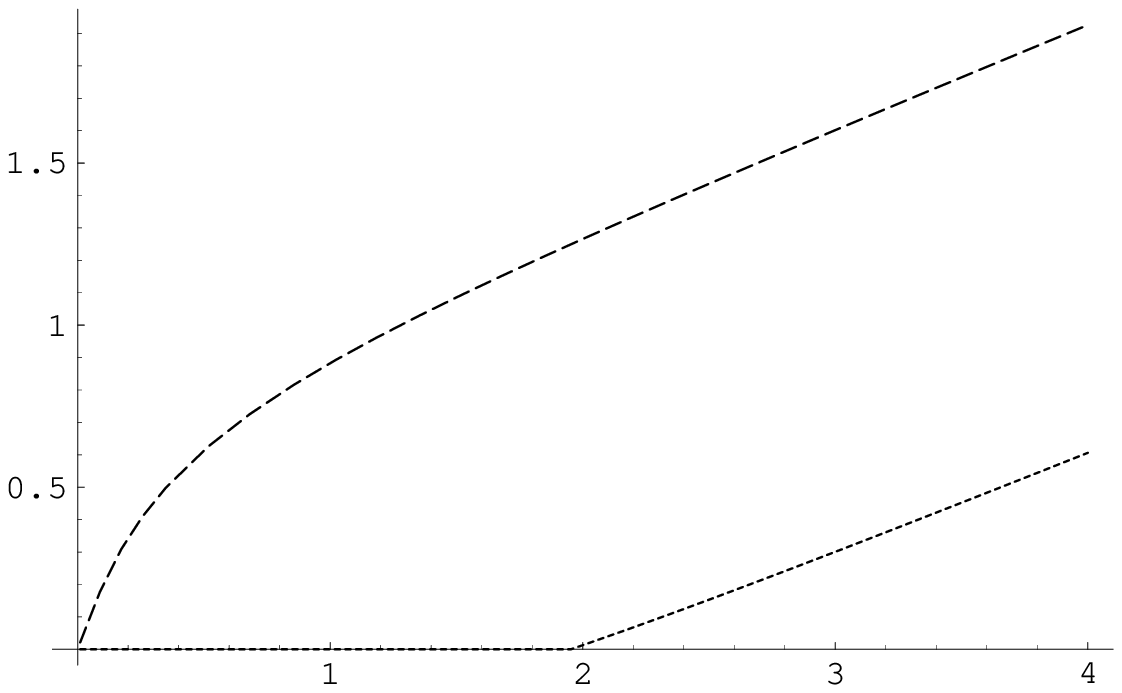}
\setlength{\unitlength}{1cm}
\begin{picture}(9,0)
\put(-1.,4.0){\makebox(0,0){\scriptsize (a)}}
\put(-4.0,3.4){\makebox(0,0){\scriptsize $\mid{\rm Im}\,\omega\mid \over m_D$}}
\put(-1.,0.2){\makebox(0,0){\scriptsize $k/m_D$}}
\put(4.8,4.0){\makebox(0,0){\scriptsize (b)}}
\put(1.8,3.4){\makebox(0,0){\scriptsize $\mid{\rm Im}\,\omega\mid \over m_D$}}
\put(4.8,0.2){\makebox(0,0){\scriptsize $k/m_D$}}
\put(10.52,4.0){\makebox(0,0){\scriptsize (c)}}
\put(7.52,3.4){\makebox(0,0){\scriptsize $\mid{\rm Im}\,\omega\mid \over m_D$}}
\put(10.52,0.2){\makebox(0,0){\scriptsize $k/m_D$}}
\end{picture}
\caption{
Dispersion relations for the unphysical $\alpha$-modes for $\theta_n=0$ and 
(a) $\xi=0.1$, (b) $\xi=1$, and (c) $\xi=10$.  Dotted lines correspond to the 
modes which depend on the sign of $f_{\rm LHP}$ (type-1). Note that in (a), (b), 
and (c) the type-1 unphysical $\alpha$-mode only exists for $k>0.184$, 
$k>0.595$, and $k>1.95$, respectively. Dashed lines correspond to modes which do 
not depend on the sign of $f_{\rm LHP}$ (type-2). Solid line in (a) is the 
isotropic result. Note that the scale changes in each plot.
}
\label{alpha:fig}
\end{figure}

A plot of the position of the unphysical $\alpha$-modes for $\xi=\{0.1,1,10\}$ 
is shown in Fig.~\ref{alpha:fig}.  Since there are two unphysical $\alpha$-modes 
we will refer to the ones that depend on the sign of $f_{LHP}$ as type-1 and 
those which don't as type-2. We can observe from this Figure that, as $\xi$ goes 
to zero, the type-1 mode (dotted line) becomes degenerate with the isotropic 
solution (solid line in Fig.~\ref{alpha:fig}a) and the type-2 mode (dashed line) 
moves off to infinity.  In the opposite limit, $\xi\gg1$,  we find that the 
type-2 mode moves towards the real axis while the threshold for existence of the 
type-1 mode moves to infinity.

Note that even in the isotropic limit there are unphysical $\alpha$-modes on the 
neighboring Riemann sheets:  one in the UHP and one in the LHP. These unphysical 
modes are related to the presence of dynamical screening of the magnetic 
interaction.  In fact, in the limit of small momentum it can be shown that these 
solutions are directly related to the small $z$ behavior of the isotropic 
transverse polarization tensor.  This can be seen by taking the small $z$ and 
$\xi$ limits of $\alpha_{UHP}$
\bqa
\lim_{z \rightarrow 0} \lim_{\xi \rightarrow 0 } {\alpha_{UHP} \over m_D^2} = 
	{i z \pi \over 4} (1 + {3 \over 8}\xi) - {1\over3}\xi + {\cal O}({\xi^2,z^2})
	\; .
	\label{smallxiza}
\eqa
\comment{mp}

The first term in (\ref{smallxiza}) comes directly from the logarithmic cut. 
Taking the isotropic limit, $\xi=0$, inserting this into the $\alpha$-mode 
equation given in Eq.~(\ref{tn0disprel}), and then taking the static limit we 
see that there is a solution at purely imaginary $\omega$
\bqa
\omega = {4 i \over \pi} {k^3 \over m_D^2 } + {\cal O}({\xi,z^2})
	\; .
\eqa
\comment{m}
Note, however, that the second term in Eq.~(\ref{smallxiza}) indicates the 
presence of unstable modes on the physical sheet for this angle of propagation 
and, as a result, the isotropic and anisotropic dispersion relations for the 
unphysical $\alpha$-modes are not trivially connected.  This is demonstrated by 
the fact that the type-1 mode (dotted lines) in Fig.~\ref{alpha:fig} do not 
extend down to $k=0$ for finite $\xi$ but instead terminate at a finite $k=k_0$. 
Below $k_0$ the type-1 mode in the LHP moves onto the physical sheet and becomes 
the unstable (anti-damped) physical $\alpha$-mode already discussed in 
Ref.~\cite{Romatschke:2003ms}.  The type-1 mode in the UHP likewise moves onto 
the physical sheet and becomes the damped physical $\alpha$-mode solution.
\comment{Is this true?: 
As a consequence below $k_0$ the dynamic screening of magnetic interactions
is determined solely in terms of type-2 unphysical $\alpha$-modes.}

\begin{figure}
\hfill
\begin{minipage}[t]{.45\linewidth}
\includegraphics[width=0.9\linewidth]{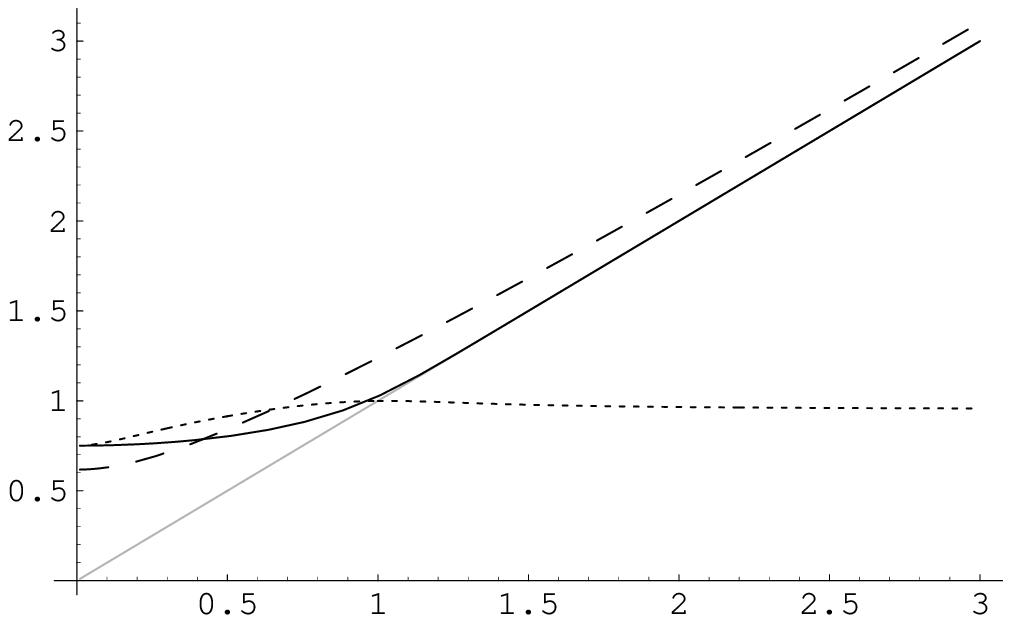}
\setlength{\unitlength}{1cm}
\begin{picture}(9,0)
\put(-0.1,3.0){\makebox(0,0){\footnotesize $\omega\over m_D$}}
\put(4,0.1){\makebox(0,0){\footnotesize $k/m_D$}}
\end{picture}
\caption{Dispersion relations for $\theta_{n}=0$, $\xi=10$: 
besides two propagating $\alpha$- and $\beta$-modes on the physical sheet 
(dashed and full lines, respectively) there is also 
an unphysical $\beta$-mode (dotted line) for which we plot only the real 
part of the pole position.  We also plot the lightcone as a 
light-gray line as a visual aide.
}
\label{disprel:fig}
\end{minipage} \hfill
\hfill
\begin{minipage}[t]{.48\linewidth}
\includegraphics[width=0.9\linewidth]{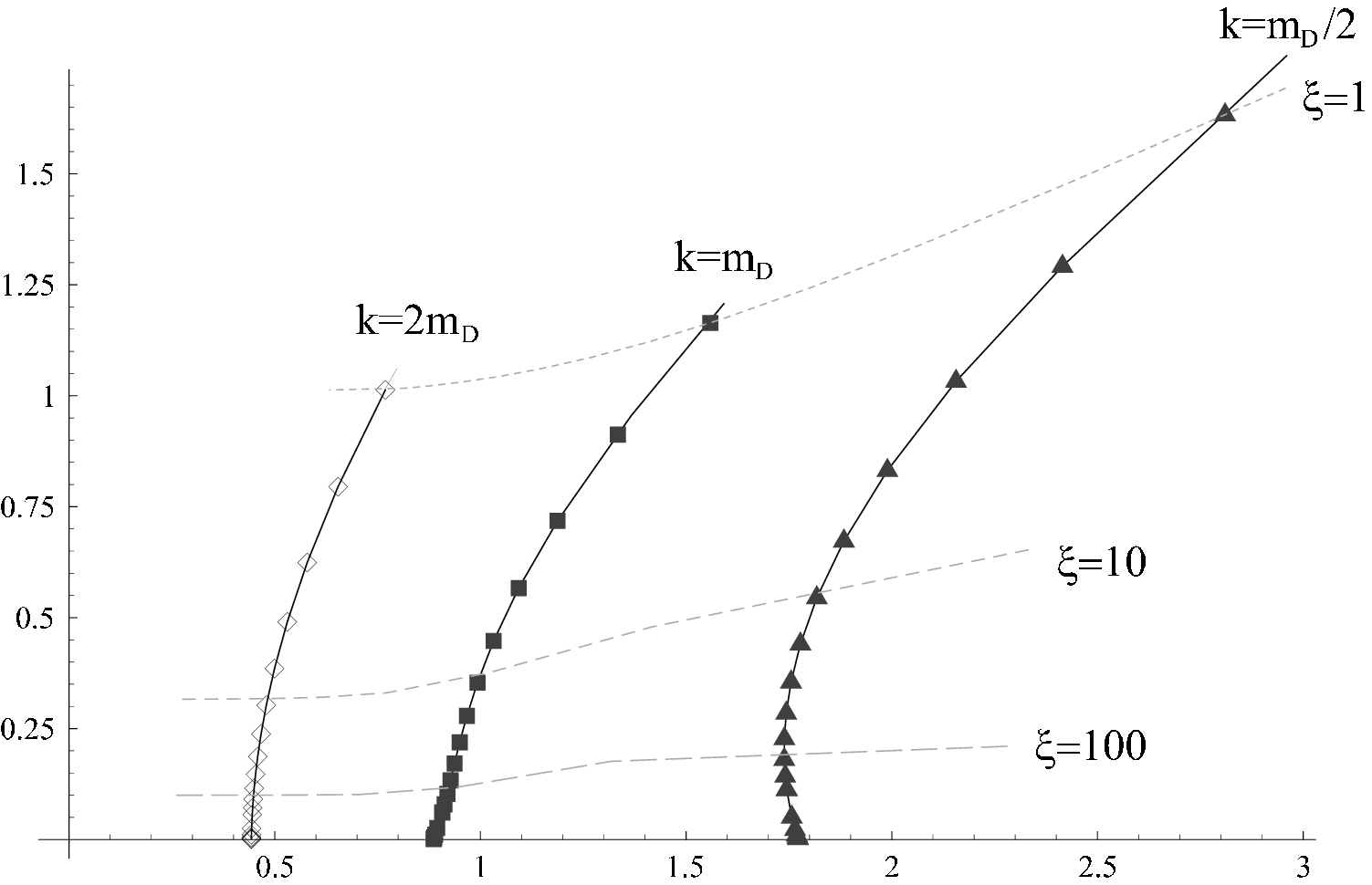}
\setlength{\unitlength}{1cm}
\begin{picture}(9,0)
\put(0.1,4.7){\makebox(0,0){\footnotesize $\mid{\rm Im}\,z\mid\over m_D$}}
\put(6.6,0.1){\makebox(0,0){\footnotesize $\mid\!{\rm Re}\,z\!\!\mid\!/m_D$}}
\end{picture}
\caption{
Parametric plot of the position of the unphysical $\beta$-mode in
the complex $z$ plane for momentum $k=\{m_D/2,m_D,2 m_D\}$ showing that 
for $k>\sim m_D$ the unphysical $\beta$-mode 
moves into the spacelike region of the real $z$-axis for large-$\xi$.
}
\label{poleposition:fig}
\end{minipage} \hfill
\end{figure}

A plot of the dispersion relations of the physical and unphysical $\beta$-modes 
for $\xi=10$ is shown in Fig.~\ref{disprel:fig} (the mirror region ${\rm 
Re}\,z<0$ is not shown). As one can see, the unphysical $\beta_{LHP/UHP}$ mode 
is lightlike for small momentum and spacelike at large momentum. As we will 
discuss below this mode is physically relevant if ${\rm Re}\,z$ is approximately 
spacelike, $({\rm Re}\,z)^2$ \raisebox{-1mm}{$\stackrel{<}{\sim}$} 1, and the 
modulus of ${\rm Im}\,z$ is small. In addition, in Fig.~\ref{poleposition:fig} 
we have plotted the position of the unphysical $\beta$-mode pole for momentum 
$k=\{m_D/2,m_D,2m_D\}$ in the complex plane for various values of $\xi$. 

From Fig.~\ref{poleposition:fig} we see that, as the anisotropy is decreased the 
${\rm Im}\,z$ of the unphysical $\beta$-mode becomes large at all momentum and 
therefore these unphysical modes will have a negligible impact on the propagator 
on the physical Riemann sheet.  However, as the anisotropy is increased with 
fixed momentum, the modulus of ${\rm Im}\,z$ decreases so that these modes can 
become relevant for large anisotropies.  From Fig.~\ref{poleposition:fig} we can 
also see that for fixed $\xi$ and decreasing $k$ that the unphysical $\beta$-
modes move to $z = \infty$ and so are unimportant based on the criteria stated
above.  However, for $k$ \raisebox{-
1mm}{$\stackrel{>}{\sim}$}  $m_D$ the unphysical $\beta$-modes have $({\rm 
Re}\,z)^2$ \raisebox{- 1mm}{$\stackrel{<}{\sim}$} 1. Therefore for large 
anisotropies and $k$ \raisebox{-1mm}{$\stackrel{>}{\sim}$} $m_D$ the effects of 
the unphysical $\beta$-modes can have an important impact on the propagator 
for spacelike modes on the physical Riemann sheet. Additionally, we see that for 
infinite $\xi$ the ${\rm Im}\,z$ of the unphysical $\beta$-mode vanishes and 
it then moves to the physical sheet as we will discuss below.

\begin{figure}
\hfill
\begin{minipage}[t]{.45\linewidth}
\includegraphics[width=0.72\linewidth]{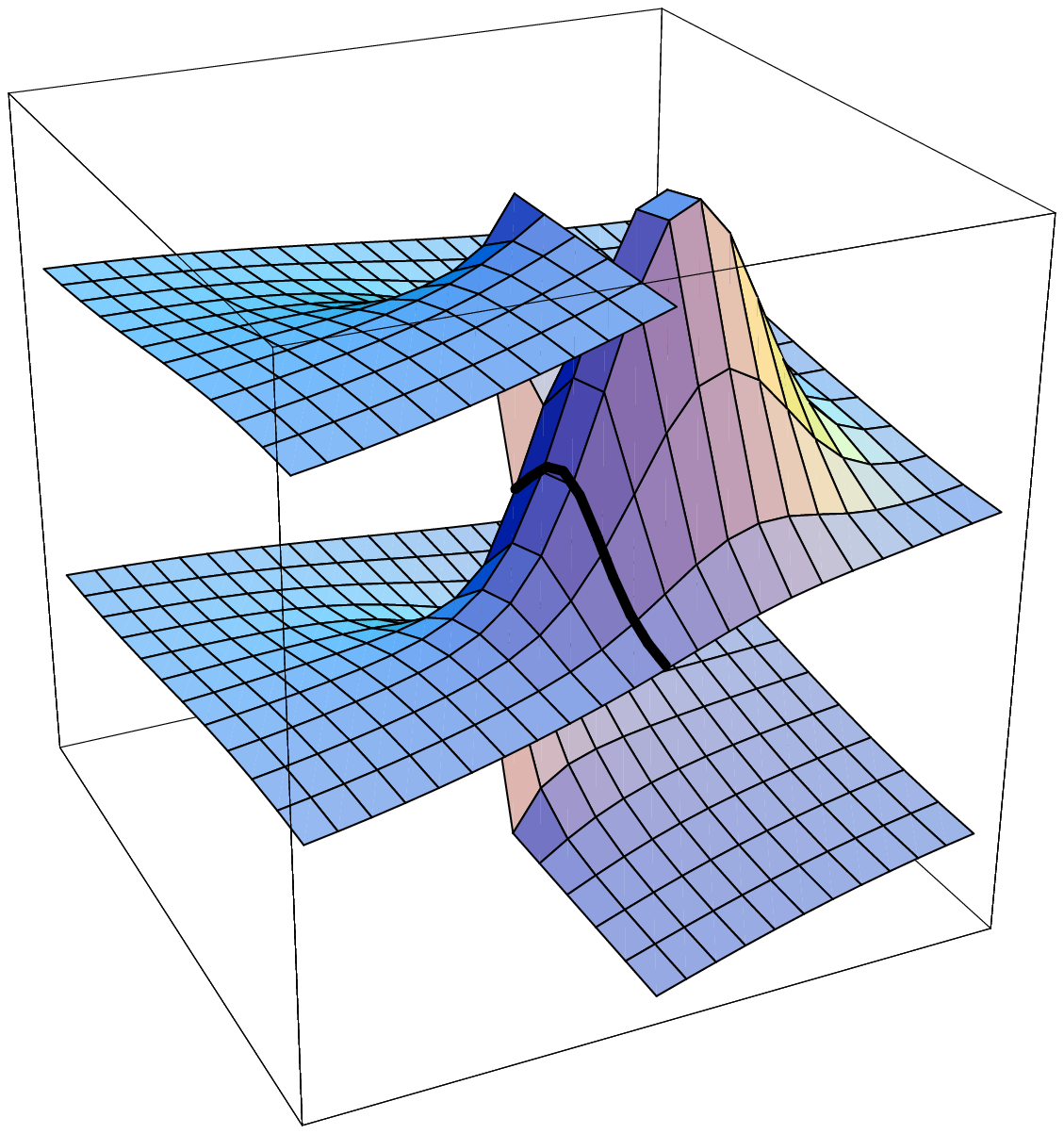}
\setlength{\unitlength}{1cm}
\begin{picture}(9,0)
\put(0.9,1.8){\makebox(0,0){\footnotesize ${\rm Re}\,z$}}
\put(5,0.8){\makebox(0,0){\footnotesize ${\rm Im}\,z$}}
\end{picture}
\caption{Sketch of the complex $z$ plane including the extension
of the logarithm to the unphysical sheet. Also shown
is how a pole in the unphysical region (mountain) has effects felt on the
physical sheet.  The black line indicates where the two sheets are joined
together.}
\label{mountain:fig}
\end{minipage} \hfill
\hfill
\begin{minipage}[t]{.45\linewidth}
\includegraphics[width=0.9\linewidth]{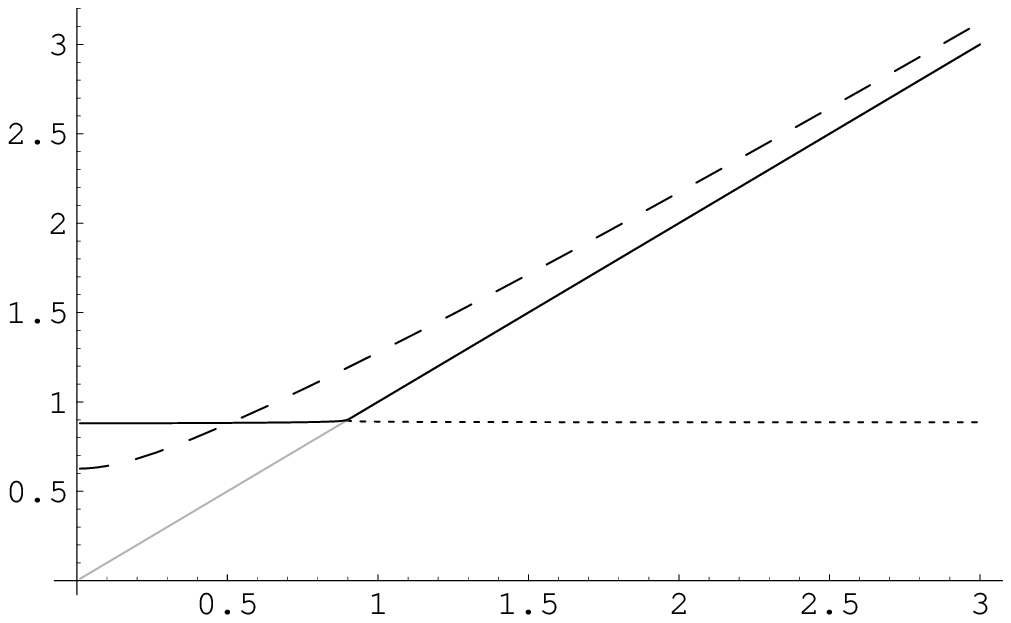}
\setlength{\unitlength}{1cm}
\begin{picture}(9,0)
\put(0.1,3){\makebox(0,0){\footnotesize ${\omega\over m_D}$}}
\put(4,0.175){\makebox(0,0){\footnotesize $k/m_D$}}
\end{picture}
\caption{
Dispersion relations for $\theta_n=0$ and $\xi=10^4$: the $\beta$-mode 
approaches its large-$\xi$ behavior.  Again the physical $\alpha$-, physical $\beta$-, 
and unphysical $\beta$-modes are indicated by dashed, solid, and dotted lines, 
respectively.
}
\label{disprel2:fig}
\end{minipage} \hfill
\end{figure}

\subsection{Mountains on spirals}

As discussed in the previous section we find that, in addition to modes on the 
physical sheet found in Ref.~\cite{Romatschke:2003ms}, for anisotropic systems 
there are also unphysical $\alpha$- and $\beta$-modes on neighboring Riemann 
sheets. But why bother about these modes, given that they do not ``live'' on the 
physical sheet? To see that there can be an effect on physical quantities, 
imagine the structure of the complex $z$ plane spanned by the different sheets 
of the logarithm in the form of a spiral staircase: the physical sheet would 
correspond to the region covered by the spiral plane from the ground floor to 
the first floor, while the unphysical sheet where the extra quasiparticle mode 
lives would correspond to the region first to second floor. 

However, since for a range of momenta the unphysical $\beta$-mode corresponds 
to an approximately spacelike pole, the propagator has 
a mountainous dent (singularity) in the spiral plane somewhere from the first to 
the second floor, with its peak the nearer the first floor the larger $\xi$ is. 
But because the mountain has a finite width, its base can be felt also below the 
second floor, especially if its peak is near the first floor (see 
Fig.~\ref{mountain:fig} for a sketch).  This is precisely the situation we 
presented in previous section.  There we showed that in the large-$\xi$ limit 
that for $k$ \raisebox{- 1mm}{$\stackrel{>}{\sim}$} $m_D$ the ${\rm Im}\,z$ of 
the unphysical $\beta$-modes becomes very small and $({\rm Re}\,z)^2$ 
\raisebox{-1mm}{$\stackrel{<}{\sim}$} 1. Therefore, we expect that the 
unphysical modes on neighboring Riemann sheets do have physical consequences for 
large anisotropies, since then their effect can extend down to the physical 
sheet. In fact, as we will discuss below, in the limit of infinite $\xi$ the 
pole moves onto the physical sheet itself. Note that the physical effect of any 
singularities existing on higher Riemann sheets ($\mid\!\!n\!\!\mid>1$) would be 
negligible since the effect of these singularities would have to extend through 
the intermediate sheets prior to getting to the physical sheet, i.e. the base of 
the mountain would have to extend all the way down to the physical sheet through 
the spacelike spirals.

\subsection{Towards large $\xi$}
\label{almostlx:sec}

For very large values of $\xi$, the dispersion relations are shown in
Fig.~\ref{disprel2:fig}. As one can see, the physical $\beta$-mode 
now hits the lightcone at the point where the unphysical
mode becomes spacelike. Indeed, when one takes the limit of 
$\xi \rightarrow \infty$, the $\beta$-mode simply becomes
\beq
1-\frac{\pi}{4} \frac{m_D^2}{\omega^2} =0 \; ,
\eeq
\comment{m}
which has the simple real and propagating solution $\omega^2=\pi m_D^2/4$. For 
any finite $\xi$ the logarithmic singularity of $\beta$ in 
Eq.~(\ref{kx0strufunc}) at the lightcone causes the physical $\beta$-mode to 
always be timelike; however, the unphysical solution exists for both timelike 
and spacelike momenta.

\section{Special Case II :  $\xi \rightarrow \infty$}
\label{lx:sec}

Another special case where one can explicitly calculate the structure functions
is when $\xi \rightarrow \infty$. In this case it has been found in
Ref. \cite{Romatschke:2003yc} that the distribution function becomes 
\beq
\lim_{\xi\rightarrow \infty} 
f_{\xi}({\bf p})
\rightarrow \delta({\bf \hat{p}}\cdot{\bf \hat n}) \int_{-\infty}^{\infty} dx\ 
f_{\rm iso}\left(p \sqrt{1+x^2}\right),
\eeq
which corresponds to the extreme anisotropic case considered by 
Arnold, Lenaghan, and Moore \cite{Arnold:2003rq}. As a consequence,
one can make use of 
this form by partially integrating Eq.~(\ref{selfenergy2}) to obtain
\beq
\Pi^{i j}(K)=2 \pi \alpha_s \int \frac{d^3 p}{(2\pi)^3} \frac{f_{\xi}%
({\bf p})}{p}%
\left[\delta^{i j}-\frac{k^i v^j+k^j v^i}{-K\cdot V-i\epsilon}+%
\frac{(-\omega^2+k^2)v^i v^j}{(-K\cdot V-i \epsilon)^2}%
\right] ,
\label{selfenergy3}
\eeq
and applying the techniques from Ref.~\cite{Arnold:2003rq} to obtain 
analytic expressions for the structure functions in the large-$\xi$ limit.
Using 
\beq
\lim_{\xi \rightarrow \infty} 2 \pi \alpha_s \int \frac{d^3 p}{(2\pi)^3} \frac{f_{\xi}%
({\bf p})}{p}= m_D^2\frac{\pi}{4} \; ,
\eeq
the structure functions are obtained using the 
contractions Eq.~(\ref{contractions}), giving
\bqa
\alpha&=&\frac{m_D^2 \pi}{4} \left[-{\rm cot}^2{\theta_n} +%
\frac{z}{\sin^2{\theta_n}}%
\left(z \pm \frac{1-z^2}{\sqrt{z+\sin{\theta_n}}\sqrt{z-\sin{\theta_n}
}%
}\right)\right] \, ,\nonumber\\
\beta&=&\frac{m_D^2 \pi}{4} z^2\left[-1\pm z%
\frac{z^2+\cos{2\theta_n}}{%
(z+\sin{\theta_n})^{3/2}(z-\sin{\theta_n})^{3/2}}\right] , \nonumber \\
\gamma&=&\frac{m_D^2 \pi}{4} \frac{1-z^2}{4\sin^2{\theta_n}}\left[%
6+2\cos{2 \theta_n}\pm z \frac{3-6 z^2-2(1+z^2)%
\cos{2\theta_n}-\cos{%
4 \theta_n}}{(z+\sin{\theta_n})^{3/2}(z-\sin{\theta_n})^{3/2}}\right] , \nonumber\\
\hat\delta&=&\frac{m_D^2 \pi}{4} \frac{\cos{\theta_n}}{\sin^2{\theta_n}}%
z\left[z\pm\frac{(1-2 z^2) 
\cos^2{\theta_n}-(1-z^2)^2}{%
(z+\sin{\theta_n})^{3/2}(z-\sin{\theta_n})^{3/2}}\right].
\label{LXstruf}
\eqa
\comment{m}
where a positive sign above corresponds to the physical sheet and a negative 
sign corresponds to the unphysical sheet. Note that the lightcone singularity at 
$z^2=1$ is not present in these structure functions any longer. In fact, all 
structure functions turn out to be purely real for $z^2>\sin^2{\theta}$ while 
there is a singularity located at $z^2=\sin^2{\theta}$. Below this lightcone-like 
structure, the imaginary parts are non-vanishing.

\subsection{Collective modes}

The dispersion relations of the collective modes are once more determined
by the zeros of Eq.~(\ref{disprel}). By conducting a Nyquist analysis
similar to the previous section, one finds
$N_{\alpha,phys}=2+2\Theta\left(-\lim_{z\rightarrow 0}\Delta_{A}(z)%
\right)$ and $N_{\Delta_G,phys}=4+2 \Theta\left(\lim_{z\rightarrow 0}%
\Delta_{G}(z)\right)$ for the physical sheet. These modes can be identified
as the standard, propagating modes (one for $\Delta_{\alpha}$ and two
for $\Delta_{G}$ together with their negative frequency equivalents) as
well as two imaginary modes (one in the UHP and one in LHP) for $\Delta_{\alpha}$ and $\Delta_{G}$,
respectively. The latter again depend only on the sign of the static
limit of the propagators, which in the limit of large $\xi$ are functions
of the momentum $k$ and the angle $\theta_n$, respectively.\footnote{%
See Ref.\cite{Romatschke:2003yc}
for a discussion of the static limit of Eq.~(\ref{LXstruf}).}

By once again extending the structure functions Eq.~(\ref{LXstruf}) to
the unphysical sheets in the LHP and UHP (the terms involving square roots
simply pick up an overall minus sign) we are also able to count
the modes there by repeating the earlier analysis. One finds
$N_{\alpha,LHP}=N_{\alpha,UHP}=\Theta\left(\lim_{z\rightarrow 0}\Delta_{A}(z)%
\right)$ corresponding to a purely imaginary mode, and
$N_{\Delta_G,LHP}=N_{\Delta_G,UHP}=2-\Theta\left(\lim_{z\rightarrow 0}%
\Delta_{G}(z)\right)$. The first two of the
latter modes are in general complex modes but
may -- in some restricted region of parameter space -- also become
purely imaginary solutions; the step function then just encodes
the fact that one of the purely imaginary solutions moves to the physical
sheet.

\begin{figure}
\includegraphics[width=0.4\linewidth]{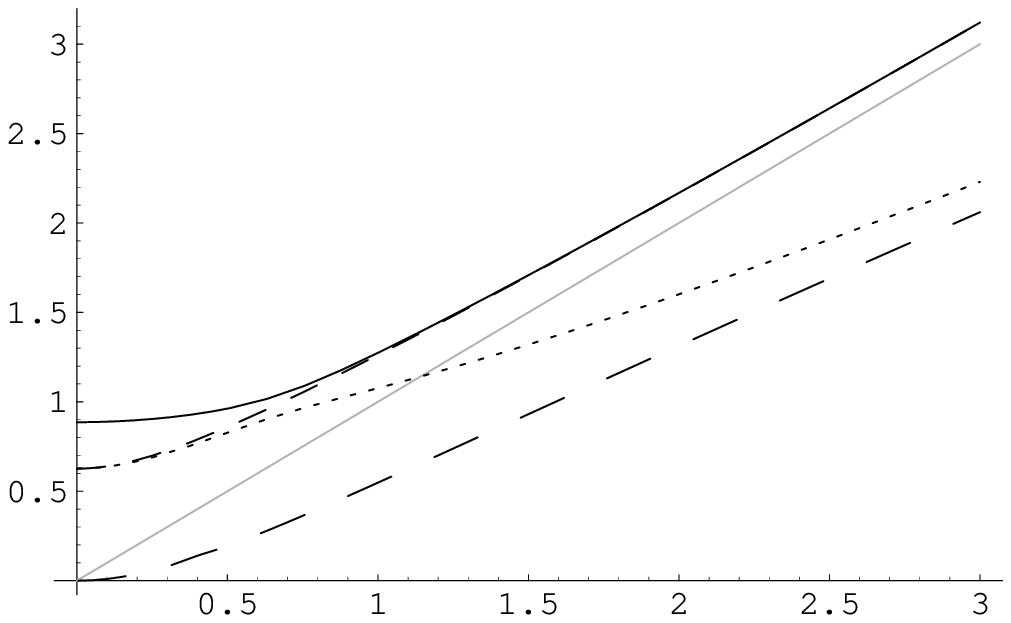}
\hspace*{1cm}
\includegraphics[width=0.4\linewidth]{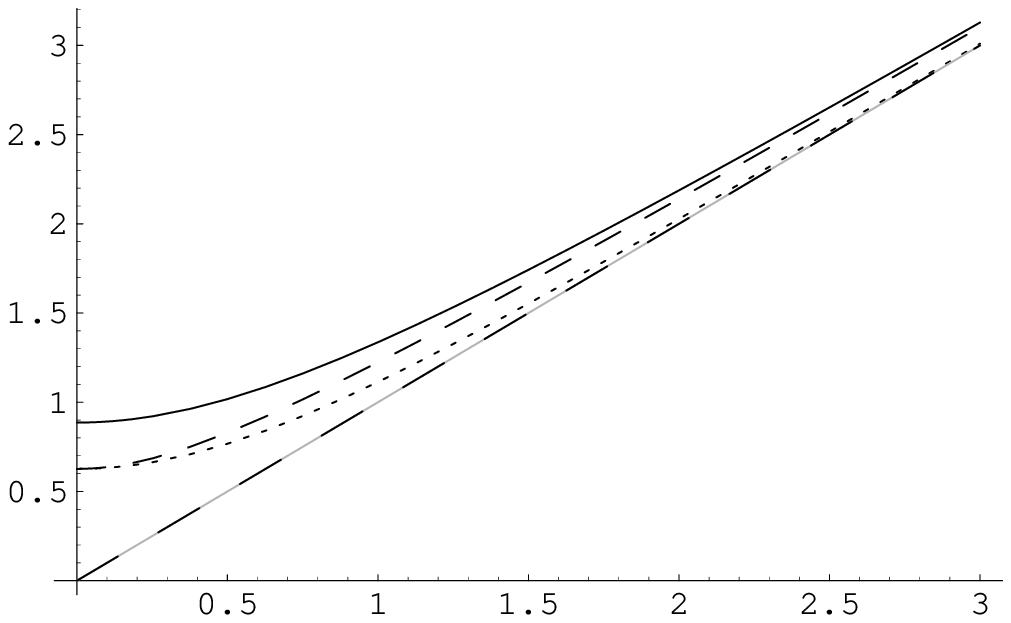}
\setlength{\unitlength}{1cm}
\begin{picture}(9,0)
\put(-3,3){\makebox(0,0){\scriptsize ${\omega\over m_D}$}}
\put(0.9,0.1){\makebox(0,0){\scriptsize $k/m_D$}}
\put(0.9,5){\makebox(0,0){\scriptsize (a)}}
\put(4.8,3){\makebox(0,0){\scriptsize ${\omega\over m_D}$}}
\put(8.7,0.1){\makebox(0,0){\scriptsize $k/m_D$}}
\put(8.7,5){\makebox(0,0){\scriptsize (b)}}

\end{picture}
\caption{
Dispersion relations for $\xi=\infty$ for (a) $\theta_n=\frac{\pi}{4}$ and 
(b) $\theta_n=\frac{\pi}{2}$. Shown are the physical $\alpha-$ mode (short 
dashed lines) and the two physical $\beta$-modes (full and dotted lines) as well 
as the unphysical $\beta$-mode (long-dashed lines), respectively. For 
$\theta_n=0$, the dispersion relations resemble those of Fig.~\ref{disprel2:fig} 
except that the mode which follows the light cone beyond $k^2=m_D \pi/4$ in 
Fig.~\ref{disprel2:fig} ceases to exist and the $\beta$-mode connects
continuously across the lightcone to the now physical spacelike mode.
}
\label{disprelLX:fig}
\end{figure}

\section{Conclusions}
\label{conc:sec}

In this paper we have extended our studied the gluon polarization tensor in an 
anisotropic system.  We extended our previous analysis to unphysical Riemann 
sheets and showed that for anisotropic distribution functions there are modes 
(singularities) in the ``spacelike region'' of the unphysical sheet which become 
physically relevant for large anisotropies.  The chief way that these modes 
affect the physics is by altering the behavior of the propagator at soft 
spacelike momenta. The behavior of the propagator in this region determines the 
rate of energy transfer from soft to hard modes and the sign of this energy 
transfer may change for anisotropic systems so that there is instead a transfer 
of energy from hard to soft modes \cite{Pitaevskii}.  Whether or not the 
presence of the modes on the unphysical Riemann sheets are responsible for this 
``anti-Landau-damping'' will be investigated in a separate paper 
\cite{Romatschke:2004el2} in which we compute heavy fermion energy loss in a 
quark-gluon plasma along the lines of Ref.~\cite{Romatschke:2004el1} in which we 
calculated the same in an anisotropic QED plasma.

In addition to these unphysical $\beta$-modes we found that for anisotropic 
distributions the unphysical $\alpha$-modes are different than in the isotropic 
case.  In the isotropic case, there are two unphysical $\alpha$-modes with one 
being in the upper half plane of the $n=1$ unphysical sheet and one in the lower 
half plane of the $n=-1$ unphysical sheet.  We showed that for finite-$\xi$ and 
small momentum these isotropic unphysical modes move onto the physical sheet and 
become the unstable modes already discussed in Ref.~\cite{Romatschke:2003ms}. 
Additionally, for finite anisotropy there are two additional unphysical 
$\alpha$-modes, again, with one being in the upper half plane and one in the 
lower half plane.  These modes would, in principle, determine the dynamic 
screening of the magnetic gluon interaction at small momentum; however, the 
instabilities in this region will dominate the physics so it's not entirely 
obvious how important these additional unphysical $\alpha$-modes are.

In closing we would like to point out that although the finite-$\xi$ analysis 
presented here was performed only for $\theta_n=0$ there are generally relevant 
unphysical $\alpha$- and $\beta$-modes for all angles of propagation.  The analysis
proceeds exactly as discussed here but the details (number of modes etc.) changes
here and there.  Additionally, the results presented here are also applicable
to anisotropic ultrarelativistic QED plasmas.

\section*{Acknowledgments}
The authors would like to that A. Rebhan for very useful discussions
and feedback.
M.S. was supported by the Austrian Science Fund Project No. M790-N08.

\bibliography{TFTSlac.bib}
\bibliographystyle{utphys}

\end{document}